\documentclass[prl,showpacs,groupedaddress,superscriptaddress,twocolumn]{revtex4}

\usepackage{fontenc,times,amsmath,amsfonts,amssymb,graphics,graphicx,epsfig,color,bbm,natbib,bm}

\begin{document}

\title{Preferred Measurements: Optimality and Stability in Quantum Parameter Estimation}

\author{Gabriel A.\ Durkin} \email{gabriel.durkin@qubit.org}
\affiliation{Quantum Laboratory , NASA Ames Research Center,  Moffett Field, California 94035, USA}

\date{\today}

 
\begin{abstract}  We explore precision in a measurement process incorporating pure probe states, unitary dynamics and complete measurements via a simple formalism. The concept of `information complement' is introduced. It undermines measurement precision and its minimization reveals the system properties at an optimal point. Maximally precise measurements can exhibit independence from the true value of the estimated parameter, but demanding this severely restricts the type of viable probe and dynamics, including the requirement that the Hamiltonian be block-diagonal in a basis of preferred measurements. The curvature of the information complement near a globally optimal point provides a new quantification of measurement stability.  \end{abstract}

\maketitle

Scientists strive  for an understanding of Nature by a physical interaction introducing correlations between observer and observed, and the process is called measurement. Fundamental limitations on the precision of any measurement exist due to the geometric distinguishability of quantum states \cite{Luis}, yet for a quantum observable there is still a classical probability distribution over measurement outcomes.  It may depend on some real-valued system parameter $\theta$ such as an interaction time or interferometric phase that has no associated Hermitian observable and is not measurable directly. Yet, the estimation of $\theta$ may be the true goal of the measurement. Inferring $\theta$ within a confidence interval (precision) $\Delta \theta$ from frequencies of measurement outcomes is a standard challenge in classical information theory with an established methodology \cite{Cover and Thomas}.  In a quantum context, the conventional approach stretches back decades \cite{Helstrom}, employing techniques from Riemannian geometry  to find precision limits via   `quantum Fisher information'  (QFI) \cite{Metrology,MetrologyII,CavesBraun94, annals} , a function of the input state and dynamics alone. It defines a precision limit for measurements but sheds little light on what measurement to use -- those proposed as optimal are typically functions of the unknown  $\theta$ \cite{SLD}.  (It is of limited utility that the parameter estimation should require prior knowledge of its true value. Adaptive techniques employing feedback have been proposed to circumvent this problem \cite{Adaptive-Lossy-Interferometry}.) 

In this letter we harness a quantum formalism for pure states emerging directly and naturally from a single result of classical information theory. The method incorporates all three instrument components --  input, dynamics (Hamiltonian) and measurement choice -- on an equal footing without recourse to QFI or the mathematical apparatus supporting it.  Important metrological results  will be confirmed in a straightforward manner along the way towards developing new conditions for optimality and stability in quantum measurements. 

A preferred measurement $\hat{M}$ has three important properties:  $(1)$ high precision, or in the many-body case, `supra-classical' precision. (By this it is meant that the precision is better for collective dynamics than is possible for component parts evolving in a separable state \cite{Metrology}.) $(2)$ If possible, $\hat{M}$ is independent of the estimated parameters \cite{annals,Phase-POVM}, and $(3)$ $\hat{M}$ is highly stable, i.e. the precision exhibits robustness against small perturbations in the state or measurement alignment. Below, these goals are given explicit mathematical expression; optimality criteria imposed on probe and dynamics.

Consider a maximal test  \cite{Peres} having outcomes labelled $k$ and an associated probability distribution $P(\theta)=\{p_k (\theta)\}$  that depends on a continuous real parameter $\theta$. The parameter estimation task involves an inference of $\theta$ from $\{ p_k (\theta)\}$. Classical Fisher information \cite{Cramer-Fisher} is defined as
\begin{equation}
\mathcal{J}(\theta) = \sum_{k} p_k (\theta) \big[\partial_{\theta} \ln p_k (\theta)\big]^2 \; , \label{F1}
\end{equation}
a measure of the information contained in the distribution $P(\theta)$ about the parameter $\theta$ \cite{Cover and Thomas}. Unlike the quantum counterpart \cite{Luis} it defines a \emph{unique} distance metric on probability space \cite{Fisher-unique-metric}. An explicit lower bound for the standard error of an unbiased estimate $\tilde{\theta}$ on the true value $\theta$ is given by the reciprocal of the Fisher information, $(\delta \tilde{\theta})^2 \geq 1 / \mathcal{J}(\theta) $,  called the Cram\'{e}r-Rao bound \cite{Cramer-Fisher}. For optimal precision one must therefore maximize $\mathcal{J}(\theta)$.

Take a complete measurement observable $\hat{M}$ with outcomes $\{m_k\}$ associated with distribution $\{p_k\}$. Apply $\hat{M}$ to a quantum system that previously evolved from a known initial state $|\psi_{0} \rangle$ under the dynamics of some time-independent Hamiltonian $\hat{H}$ for time $\theta$.  The Schr\"{o}dinger equation governs the dynamics $ i \partial_{\theta}  | \psi_{\theta} \rangle = \hat{H} | \psi_{\theta} \rangle $,
(where $\hbar=1$) and the time evolution is explicitly $ | \psi_{\theta} \rangle = \exp\{- i \hat{H} \theta\} | \psi_{0} \rangle \; .$ Writing the spectral decomposition of the  measurement as $ \hat{M} =  \sum_{k} m_k | k \rangle \langle k | $ then  complex amplitudes $\langle k | \psi_{\theta} \rangle = r_k \exp\{i \phi_k\}$
give probabilities $ p_k = \langle k | \psi_{\theta} \rangle \langle \psi_{\theta} |  k \rangle = r_{k}^{2} \; , \label{probs}$ where $\{p_k, r_k \} \in [0,1]$ and $\phi_k \in [0,2 \pi )$ are all real-valued functions of $\theta$. Replacing $p_k$ with $r_k^2$ in Eq.\eqref{F1} gives:
\begin{equation}
\mathcal{J}(\theta)= 4\sum_k \dot{r}_{k}^{2}  \; ,\label{FI2}
\end{equation}
which we can now use to find an operator expression for the classical Fisher information. Differentiating $r_k^2$ gives
\begin{equation}
2 \dot{r}_k r_{k}= \langle k | \psi_{\theta} \rangle \langle  \dot{\psi}_{\theta} |  k \rangle + \langle k | \dot{\psi}_{\theta} \rangle \langle \psi_{\theta} |  k \rangle  \; . \label{diff}
\end{equation}
Now, $ \langle k | \dot{\psi}_{\theta} \rangle  = \partial_{\theta} (r_k e^{i \phi_k})= e^{i \phi_k} (\dot{r}+i r \dot{\phi}_k) =  | \langle k | \dot{\psi}_{\theta} \rangle| e^{i ( \phi_k + \tau_k)} $ where we define a velocity vector with radial and transverse components $\{ \dot{r}_k, r_k \dot{\phi}_k \}$, and inclination:
\begin{equation}
\tau_k  = \tan^{-1}( r_k \dot{\phi}_k / \dot{r}_k) = \arg  \langle k | \dot{\psi}_{\theta} \rangle  - \arg \langle k | \psi_{\theta} \rangle \; .
\end{equation}
\begin{figure}
\includegraphics[width=0.9in]{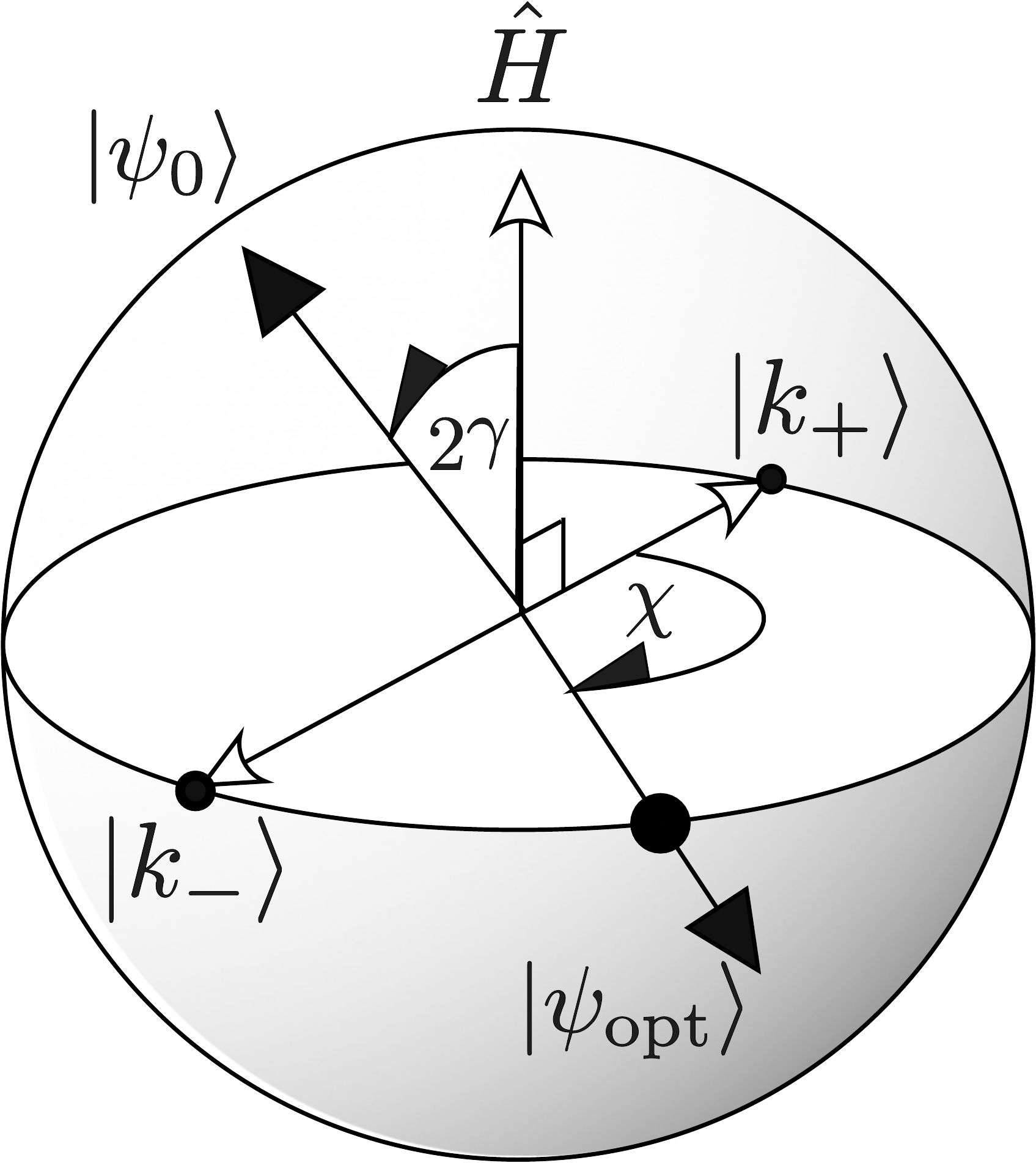}
\caption{\small{An optimal measurement scheme for a Hamiltonian $\hat{H}$ spanning any number of dimensions may be restricted to the qubit subspace of its extremal eigenvectors $| \lambda_{\uparrow}\rangle$ and $| \lambda_{\downarrow} \rangle$.  (Other measurement elements  $\in \{|k\rangle\}$ span an orthogonal subspace.)  On the Bloch sphere: the diagonal basis of $\hat{H}$ defines the $z$-axis, and the optimal measurement projectors $| k_{\pm} \rangle$ the $x$-axis. An optimal probe state $| \psi_{\text{opt}}\rangle$ lies anywhere in the equatorial plane.}} \label{sphere}
\end{figure}
Eq.\eqref{diff} yields  $\dot{r}_k  = \cos \tau_k  | \langle k | \dot{\psi}_{\theta} \rangle|  $. From the Schr\"{o}dinger equation we also have $ i \langle k | \dot{\psi}_{\theta} \rangle =  \langle k | \hat{H} |  \psi_{\theta} \rangle $. Substituting this expression and squaring gives $
\dot{r}_{k}^{2}  = \cos^2 \tau_k    \langle  \psi_{\theta} | \hat{H} | k  \rangle  \langle  k | \hat{H} | \psi_{\theta} \rangle . $ Summing over all outcomes `$k$' gives the Fisher information using Eq.\eqref{FI2}: $\mathcal{J}(\theta) = 4 \sum_k   \cos^2 \tau_k   \langle  \psi_{\theta} | \hat{H} | k  \rangle   
  \langle  k | \hat{H} | \psi_{\theta} \rangle $. Thus we can define a non-linear, positive and hermitian operator $\hat{F}_{\theta}$, diagonal in the measurement basis:
\begin{equation}
\hat{F}_{\theta} = 4 \sum_k \cos^2  \tau_k | k \rangle \langle k | = 4 \sum_k c_{\psi,k}  | k \rangle \langle k |   \: , \label{FOp}
\end{equation}
such that the Fisher information is then
\begin{equation}
\mathcal{J}(\theta) =   \langle  \psi_{\theta} | \hat{H} \hat{F}_{\theta}  \hat{H} | \psi_{\theta} \rangle = \langle  \psi_{0} | \hat{H} \hat{\Phi}_{\theta}  \hat{H} | \psi_{0} \rangle,   \label{FisherInfo}
\end{equation}
where $\hat{\Phi}_{\theta} = e^{i \hat{H} \theta}   \hat{F}_{\theta} e^{- i \hat{H} \theta}  =  4 \sum_k c_{\psi,k}  | k'  \rangle \langle k' | $ is the unitarily transformed operator. Due to the non-linear nature of $\hat{F}_\theta$, a basis transformation $| k \rangle \mapsto | k' \rangle$ gives a different result: $\hat{F}_\theta \mapsto \hat{F}_\theta' = 4 \sum_k c_{\psi,k'}  | k'  \rangle \langle k' | \; $, not equivalent to $\hat{\Phi}_\theta$, since $c_{\psi,k'}  \neq c_{\psi,k} $ generally.  This explicit definition of $\hat{F}_{\theta}$ will be useful in the qubit optimization to come.

\emph{Fixed Probe Optimization}: Now let us establish an upper bound for $\mathcal{J}(\theta)$ in terms of a fixed input and dynamics, but varying the measurement. The completeness of the measurement basis provides a resolution of the identity  $ \sum_k | k \rangle \langle k | = \mathbbm{1} $,  and therefore, with $\cos^2 \tau_k = 1- \sin^2 \tau_k $, Eq.\eqref{FisherInfo} becomes:
\begin{align}
\frac{\mathcal{J}(\theta) }{4} & = \langle \hat{H}^2 \rangle - \sum_k \sin^2  \tau_k \:  \langle  \psi_{\theta} | \hat{H} | k  \rangle   
  \langle  k | \hat{H} | \psi_{\theta} \rangle  \nonumber \\
   & =  \langle \hat{H}^2 \rangle - \sum_k ( r_k \dot{\phi}_k )^2 =  \langle \hat{H}^2 \rangle - \mathcal{K}(\theta) \; ,    \label{Fisher1}
\end{align} 
where we define the `\emph{information complement}' :
\begin{equation}
\mathcal{K}(\theta) =  \sum_k ( r_k \dot{\phi}_k )^2  = \sum_k p_k \dot{\phi}_k ^2    = \langle \dot{\varphi}^2 \rangle_c
\end{equation}
 a non-negative functional of the probe, measurement and Hamiltonian that can only reduce $\mathcal{J}$. Here $\dot{\varphi}$ is a classical random variable taking values from the set $\{ \dot{\phi}_k \}$ and the $c$ subscript denotes the expectation value is classical. We now look for a basis $\{ | k \rangle \}$ and associated set $\{ r_k, \dot{\phi}_k\}$ that minimizes $\mathcal{K}$ for a fixed input $|\psi_0\rangle$.  First,  a new description of the Hamiltonian expectation value is needed:
 \begin{align}
 \langle \hat{H} \rangle & = i \langle \psi | \dot{\psi} \rangle = i \sum_k \langle \psi |  k \rangle \langle k | \dot{\psi} \rangle = i \sum_k ( r_k \dot{r}_k + i r_k^2 \dot{\phi}_k ) \nonumber \\
 &  = - \langle \dot{\varphi} \rangle_c + \frac{i}{2} \partial_\theta \bigg( \sum_k  r_k^2   \bigg) = -  \langle \dot{\varphi} \rangle_c  \label{Hphi}
\end{align}
Therefore, $\mathcal{K}-  \langle \hat{H} \rangle^2 = \Delta_c^2 \dot{\varphi} \geq 0$. Comparing Eq.\eqref{Fisher1} it follows directly that $
\mathcal{J} \leq 4 \Delta^2 \hat{H}
$ for a fixed input  $| \psi_0 \rangle$. (This was derived by a different method in \cite{annals}.) The bound is saturated by a particular qubit input, as we will now show.

\emph{Optimizing for a Qubit:}  Probes that are eigenstate of  $\hat{H}$  give  $\dot{r}_k \mapsto 0 $ and $\mathcal{J} = 0 $ from Eq.\eqref{FI2} because any eigenstate $| \lambda \rangle$ of $\hat{H}$ only gains a phase during its evolution. Thus $r_k (\theta)= |\langle k | e^{i \hat{H} \theta}| \lambda \rangle | =  |\langle k | e^{i \lambda \theta}| \lambda \rangle | =  |\langle k | \lambda \rangle | = r_k(0)$, and $\dot{r}_k = 0$.
For optimality over all $| \psi_0 \rangle$ and $\{ | k \rangle \}$ the input state must thus be a \emph{superposition} of at least two Hamiltonian eigenvectors, $
 | \psi_0 \rangle \mapsto \cos \gamma  |  \lambda_{1} \rangle + e^{i \chi} \sin \gamma | \lambda_2 \rangle $. Choose such a qubit probe and a measurement basis  $\{ | k_1 \rangle, | k_2 \rangle\}$ spanning the same  $\mathbbm{C}^2$ as $| \lambda_{1,2} \rangle: 
| k_1 \rangle = \cos \alpha | \lambda_1 \rangle  + \sin  \alpha | \lambda_2 \rangle, \;
| k_2 \rangle =   -  \sin \alpha | \lambda_1 \rangle + \cos \alpha | \lambda_2 \rangle$. This is a two-dimensional subspace of the full Hilbert space supporting $\hat{H}$. Here it has been chosen that $\{ | k_{1,2} \rangle$ defines the $x$ axis on the Bloch sphere of FIG.\ref{sphere}, hence Im$\langle k | \lambda \rangle$=0. By confining $| k_{1,2} \rangle$ to the span of  $ | \lambda_{1,2} \rangle$ then (using the fact it is diagonal in the measurement basis) one restricts interest to the component of   $\hat{\Phi}_\theta$ within this qubit space too:
\begin{align}
\hat{\Phi}_\theta =  & \; e^{i \hat{H} \theta} \{  c_{\psi,k1}  | k_1 \rangle \langle k_1  | +  c_{\psi,k2} | k_2 \rangle \langle k_2  | + \dots \nonumber \} e^{- i \hat{H} \theta}  \nonumber \\
=  & \; \;   c_{\psi,k1}  \left(
\begin{array}{ll}
c^2  & e^{-i ( \lambda_2-\lambda_1) \theta } s c \\
e^{+i  (\lambda_2-\lambda_1) \theta } s c & s^2
\end{array}
\right) \nonumber \\ + & \;  c_{\psi,k2}  \left(
\begin{array}{ll}
  s^2  &  - e^{-i  (\lambda_2-\lambda_1) \theta) } s c \\
- e^{+ i  (\lambda_2-\lambda_1) \theta)}  s c & c^2 
\end{array}
\right) +  \dots \nonumber
\end{align}
where $ c $  ($s$) is $ \cos \alpha$  ($ \sin \alpha$).  We ignore elements of $\hat{\Phi}_\theta$ that project onto the remaining Hilbert space, orthogonal to   $\{ | \lambda_1 \rangle, | \lambda_2 \rangle\}$. Defining $ \beta   =\chi - (\lambda_2 - \lambda_1) \theta  $, then angles $\{ \alpha, \beta, \gamma \}$ give an expectation value $\langle \psi_0 | \hat{H} \hat{\Phi}_{\theta}  \hat{H} | \psi_0  \rangle$:
\begin{widetext}
\begin{equation} \! \mathcal{J}(\alpha, \beta, \gamma)\! = \! \frac{- 4  \left(\lambda _1-\lambda
   _2\right){}^2 s^2[2 \alpha ] s^2[2 \gamma ] s^2\left[ \beta \right]}{\left(c[2 (\alpha -\gamma )]+c [2 (\alpha +\gamma )]+2 c \left[ \beta \right] s [2 \alpha ] s [2 \gamma ]-2\right) \! \left( c [2 (\alpha -\gamma )]+c [2 (\alpha +\gamma )]+2 c \left[ \beta \right] s [2 \alpha ] s [2 \gamma ]+2 \right)}
\end{equation}
\end{widetext}
writing sin as `$s$' and cos as `$c$'.  $\mathcal{J}$ is optimized by angles $\{ \alpha, \gamma \} \mapsto \pi / 4$, independent of the value of $\beta$ and giving a saturable bound: $ \langle \psi_0 | \hat{H} \hat{\Phi}_{\theta}  \hat{H} | \psi_0  \rangle \leq  \left(\lambda _1-\lambda_2\right){}^2 $. This is the upper bound on the Fisher information for any superposition of two eigenstates of the Hamiltonian. It is saturated by a probe state $| \psi_{\text{opt}} \rangle =  (| \lambda_1 \rangle + e^{ i \chi } | \lambda_2 \rangle) / \sqrt{2} $, where $\chi \in [0,2 \pi)$, see FIG.\ref{sphere}. The result $\alpha \mapsto \pi/4 $ dictates an optimal measurement scheme with components:
\begin{equation}
| k_{\pm} \rangle =  (| \lambda_1 \rangle \pm | \lambda_2 \rangle) / \sqrt{2} \; ,  \label{best-measurement}
\end{equation}
also in FIG.\ref{sphere}. Other basis elements $\in \{ | k \rangle \}$ span an orthogonal subspace. It is significant that the optimal measurement is independent of $\beta$, and hence the estimated parameter $\theta$.

\emph{Generalization to Higher Dimensions:} The above result shows that for a given $\hat{H}$, the maximal Fisher information is  bounded from below  by $\left(\lambda _\uparrow - \lambda
   _\downarrow \right)^2 = || \hat{H} ||^2$ where $\lambda_\uparrow$ ($\lambda_\downarrow$) is the max (min) eigenvalue of $\hat{H}$, and $|| \hat{H}||=  \left(\lambda _\uparrow - \lambda_\downarrow \right)$ is the operator seminorm of the Hamiltonian \cite{MetrologyII}.  The variance has the seminorm as an upper bound: $ || \hat{H} ||^2 \geq 4 \Delta^2 \hat{H}$, creating a bridge between the qubit result with that for a fixed $| \psi_0 \rangle$ in a higher dimensional space. We saw for a fixed $|\psi_{0}\rangle$ that $\mathcal{J} \leq 4 \Delta^2 \hat{H}$. Therefore the qubit maximum variance state must be the universally optimal state over the full Hilbert space; it saturates the variance bound. Concisely: 
\begin{equation} \label{seminorm}
  \begin{array}{c}
    \text{max} \\ 
    | \psi_0 \rangle, \{ | k \rangle \} \\ 
  \end{array} \mathcal{J} =  || \hat{H} ||^2 \; .
\end{equation}
(This bound has been discussed previously in terms of quantum Fisher information \cite{MetrologyII}.) A corollary of Eq.\eqref{seminorm} is that no greater number of superposed energy eigenstates can be used as an input to improve on the Fisher information provided by the (qubit) maximum variance state $(|\lambda_{\uparrow}  \rangle + e^{i \chi} | \lambda_{\downarrow} \rangle )/ \sqrt{2}$. The optimal measurement set can be chosen as the  one with two elements straddling the qubit subspace of extremal eigenvalues in Eq.\eqref{best-measurement}. Importantly,  it retains independence from the true value of $\theta$.

\emph{Optimal Measurement Criterion}: There may certainly exist more than one optimal measurement set for a given input, and we can define an optimal measurement as one that satisfies $\mathcal{K}  =  \langle \hat{H} \rangle^2 $, i.e. for measurements saturating  $\mathcal{J} \leq 4 \Delta^2 \hat{H}$:
\begin{equation}
\Delta_c^2 \dot{\varphi} = \mathcal{K} -  \langle \hat{H} \rangle^2 =\sum_k  r_k^2  \big(\dot{\phi}_k - \langle \dot{\varphi} \rangle_c \big) ^2 \mapsto 0
\end{equation}
Using Eq.\eqref{Hphi},  for optimality $\dot{\phi}_k =  \langle \dot{\varphi} \rangle_c  =  - \langle \hat{H} \rangle, \forall k$. This is equivalent to a  condition presented in \cite{CavesBraun94},  $ \text{Im} \langle \psi_\theta | k \rangle \langle k | \psi_{\perp} \rangle = 0, \forall k$, where $ | \psi_{\perp} \rangle = |\dot{\psi} \rangle - \langle \psi_{\theta} | \dot{\psi} \rangle | \psi_{\theta} \rangle$ is the component orthogonal to $|\psi_{\theta} \rangle$ in the qubit space spanned by $\{ | \psi_\theta \rangle, |  \dot{\psi} \rangle \}$. Measurements that are superpositions of $\{|\psi_{\theta} \rangle,  | \psi_{\perp} \rangle \}$, e.g. those of \cite{SLD}, are generally only \emph{instantaneously} optimal -- as the state evolves, precision decreases. Demanding parameter independent optimality means $\dot{\phi}_k = - \langle \hat{H} \rangle$ at \emph{all} times, i.e.
\begin{equation}\label{cond1}
\phi_k= - \langle \hat{H} \rangle \theta + \zeta_k \;  \; , \; \; \; \forall \;  k, \theta
\end{equation}
Enforcing this condition will limit the viable probes, measurements and dynamics. Starting with $e^{i \phi_k} (\dot{r}+i r \dot{\phi}_k) = \langle k | \dot{\psi} \rangle = - i \langle k | \hat{H} | \psi \rangle$  from just after Eq.\eqref{FI2} and substituting Eq.\eqref{cond1} restricts the re-zeroed Hamiltonian, $\tilde{H} = \hat{H} - \langle \hat{H} \rangle \hat{\mathbb{I}}$ :
\begin{equation}
\tilde{H}^{(I)} | \psi_{\theta} \rangle = i | \dot{\psi} \rangle\;  \; , \; \;  \tilde{H}^{(R)}  | \psi_{\theta} \rangle = 0 . \label{idealcond}
\end{equation}
Here we have chosen an optimal $| k \rangle$ basis, in which $\tilde{H}$ has a real (imaginary) part denoted by $R$ $(I)$. Only the imaginary part determines dynamics \cite{Kimura}, and $ | \psi_{\theta} \rangle$ is confined to the null space of $\tilde{H}^{(R)} $. That $ \tilde{H}^{(R)} \exp\{-i \tilde{H}^{(I)} \delta \theta\} | \psi_{0} \rangle = 0 $ implies $[\tilde{H}^{(R)},\tilde{H}^{(I)}  ] = 0$. Therefore, in the optimal basis the Hamiltonian is block diagonal: $\tilde{H} = \tilde{H}^{(R)}  \oplus \tilde{H}^{(I)} $. Note the null space of $\tilde{H}^{(R)}$ needs to be at least two-dimensional for $ | \psi_{0} \rangle$ to evolve at all -- we have seen that maximum precision is possible in just such a qubit space. (Then  $\tilde{H}^{(I)}$ must be proportional to the Pauli $\sigma_y$.) Complete measurements like $\hat{M}$ are \emph{not} covariant; see \cite{annals,Phase-POVM,Berry-Wiseman} for a discussion of over-complete covariant measurements.

\emph{Stability at Optimal Point:} Fulfilling the conditions above may indeed give maximum precision but what if small deviations in the measurement orientation lead to a dramatic reduction in the Fisher information? If the measurement basis is rotated slightly by $| k \rangle \mapsto \exp\{- i \hat{h} \delta \omega  \} | k \rangle$  then the overlap becomes
$\langle k | \psi_{\theta} \rangle \mapsto \langle k | e^{i \hat{h} \delta \omega } e^{- i \hat{H} \delta \theta}  |  \psi_{\theta} \rangle$,
combining dynamics for measurement  drift and state evolution. The information complement is now a function of two variables, $\mathcal{K}(\theta,\omega)$. At a turning point we have $\partial_{\theta} \mathcal{K} = \partial_{\omega}  \mathcal{K}  = 0$, but this doesn't indicate much. However, at the \emph{global} minimum ($gm$) for parameter-independent evolution Eq.\eqref{cond1} applies and therefore  $\ddot{\phi}_{k} = - \partial_{\theta} \langle \hat{H} \rangle = 0, \forall k$. Here derivatives are $\partial_{\theta} \mathcal{K}=  \partial^{2}_{\theta} \mathcal{K}= 0 $, meaning that to enforce the optimality condition for all $\theta$ values confirms a zero curvature of $\mathcal{J}$ in the direction of evolution. Denoting derivatives of $\theta$ by a dot and of $\omega$ by a dash, the other second order derivatives for parameter-independent evolution are:
\begin{align}
\! \! \left. \frac{\partial^2 \mathcal{K}}{\partial \omega \partial \theta} \right|_{gm} \! \! \! \! \! \! \! \!   & = - 2 \langle \hat{H} \rangle  \sum_k  \dot{p}_k \dot{\phi}'_k  \; , \nonumber \\
\! \! \left. \frac{\partial^2 \mathcal{K}}{\partial \omega^2} \right|_{gm} \! \! \! \!  \! \! \! \!   & = 2 \langle( \dot{\varphi}')^2 \rangle_c \! - \! 2 \langle \hat{H} \rangle \bigg[ \langle \dot{\varphi}''\rangle_c \! + \! 2 \! \sum_k \! p'_k \dot{\phi}'_k \!  \bigg]
\end{align}
Probes returning $ \langle \hat{H} \rangle = 0$ produce a minimum in $\mathcal{K}$ for deviations $\delta \omega$ from the optimal measurements -- then $\partial_{\omega} \mathcal{K}= 0$ and $\partial^{2}_{\omega} \mathcal{K} \geq 0$. (Precision is optimal in both evolution and drift variables.) Ideally $\langle \hat{H} \rangle =  \partial^{2}_{\omega \theta} \phi_k = 0 , \: \forall k $, then optimal measurements are perfectly stable, i.e. $\mathcal{K} $ has a zero Hessian. Well, $\langle \hat{H} \rangle = 0$ is easily obtained by a recalibration of the zero energy point, $\hat{H}\mapsto \tilde{H}$ without changing the physics of the system. However, the mixed second derivative of $\phi_k$ cannot be zero if drift dynamics rotate the measurement basis closer to the eigenbasis of $\hat{H}$. (Parameter information becomes zero when measurements are  $\hat{H}$ eigenstates.) The ideal scenario of Eqs.\eqref{idealcond} produces $\text{Tr}[\hat{M} \Tilde{H}^{(I)}] = 0$ and maintaining this under drift $e^{- i \omega \hat{h}} \hat{M} e^{ i \omega \hat{h}}$ implies an orthogonality condition:
\begin{equation}
\text{Tr}\left( \Tilde{H}^{(I)} \: [\hat{M} ,\hat{h}] \right) = 0
\end{equation}
\begin{figure}\centering
\includegraphics[width=3.65in]{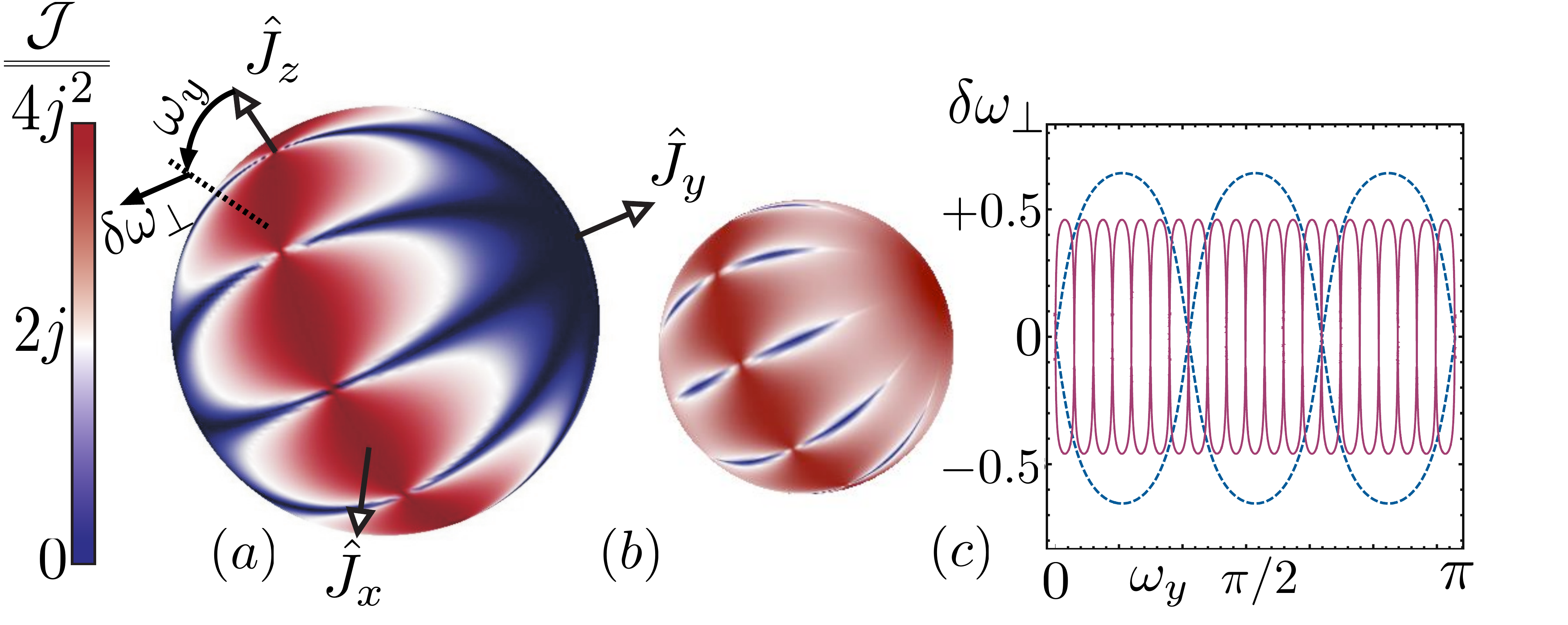}
\caption{\small{For a spin Hamiltonian $\hat{H} \mapsto \hat{J}_y$ the maximum variance state $|\psi_0\rangle=( |j,+j\rangle_y + \exp \{i \chi\} | j, - j \rangle_y ) / \sqrt{2}$ (sometimes called a NOON state \cite{NOON-both,Durkin-Dowling}) yields greatest precision: $\mathcal{J}= || \hat{J}_y||^2= 4j^2$. Eigen-equations are $\hat{J}^2 | j,m \rangle_i = j(j+1)| j,m \rangle_i$ and  $\hat{J}_i | j,m \rangle_i = m | j,m \rangle_i$ for $i \in \{x,y,z \}$. The optimal measurement set is not unique; one is given by Eq.\eqref{best-measurement} (with $\lambda_{1,2} \mapsto \pm j$) but others include $\hat{J}_x$. Spherical surfaces $(a)$ and $(b)$ are coloured by precision: Fisher information for rotational drift $|k \rangle \mapsto \exp \{i \hat{J}_z \omega_z \} \exp \{i \hat{J}_y \omega_y\} |k \rangle$ of optimal measurements away from $\hat{J}_x$  in $(a)$ or from the set of Eq.\eqref{best-measurement} in $(b)$. Blue regions correspond to precision below the classical bound (associated with $\mathcal{J}_{\text{max}}$ for $n=2j$  individual spin $1/2$ particles, $\mathcal{J} \leq 2j$). Red regions indicate supra-classical precision $2j  < \mathcal{J} \leq 4j^2$ where quantum correlations contribute to the precision \cite{Heisenberg-Limit}. The clamshell structure of  $(a)$ is characteristic of the measurement landscape at all $j$ with the red zones or "hotspots" numbering $4j$ around the equator, each decreasing in size as $j$ increases. In $(a)$ precision is highest and curvature is zero along the equator, showing that measurements $\cos \xi \hat{J}_x + \sin \xi \hat{J}_z$ are optimal and  parameter independent for dynamics $\hat{J}_y$ . However, curvature is non-zero along lines of longitude and dependent on drift angles $\omega_{y,z}$ so $|\psi_0\rangle$ may offer greatest precision but low stability. Precision contours in $(c)$ mark the classical precision boundary for NOON states of spin $j=3/2$ (dashed) and $21/2$ (unbroken). The angle $\delta \omega_{\perp}$ indicates how much transverse drift can be tolerated while maintaining supra-classical precision.}} \label{NOONBall}
\end{figure}

\emph{Summary and Outlook}: The formalism we have developed incorporates all aspects of quantum parameter estimation explicitly; probe $| \psi_{0} \rangle $, dynamics $\hat{H}$, and measurement, $\{ | k \rangle \}$, clarifying how precision is determined by the interplay of all three. The information complement $\mathcal{K}$ was introduced, a measurement dependent functional that undermines the Fisher information.  At  the global optimum for parameter-independent measurement, phases $\phi_k =\arg \langle k | \psi_{\theta} \rangle$ vary linearly with the parameter $\theta$ in proportion to their average energy, restricting dynamics to imaginary Hamiltonians in the optimal measurement basis. 

Greatest possible precision within the Hilbert space spanned by the Hamiltonian exists in the qubit subspace of the maximal variance input state and ultimate precision over all probes and measurements is completely defined by the extremal energy eigenvalues. No additional dynamical structure is relevant, nor is the dimension of the Hilbert space. The most precise measurement is also parameter free.

Stability of optimal measurements was quantified in terms of the curvature of $\mathcal{K}$ in the vicinity of its global minimum. For dynamics $\hat{h}$ causing unitary drift of measurement orientation the curvature indicates whether the apparatus is of pragmatic utility, quantifying its immunity to alignment errors at the optimal setting. States with zero average energy $\langle \hat{H}\rangle$ are associated with optimal parameter-independent measurements that become suboptimal given measurement drift, at a rate determined by the magnitude of mixed second order derivatives of phases $\phi_k$. See FIG.\ref{NOONBall} for an illustration of the precision terrain for spin states proposed previously for parameter estimation.

In future, it may prove fruitful to develop the approach presented here to incorporate evolution governed by completely positive maps and generalised measurements. This work was carried out under a contract with Mission Critical Technologies at NASA Ames Research Center. The author thanks Hugo Cable, Gen Kimura and Vadim Smelyanskiy for useful discussions

\bibliographystyle{apsrev}

\begin{thebibliography}{31}
\bibitem{SLD}\bibinfo{note}{One optimal measurement is formed by the eigenvector set for the density matrix's symmetric logarithmic derivative,  taken with respect to the estimated parameter. See \cite{CavesBraun94, annals}.}
\bibitem{Kimura}\bibinfo{note}{Generally, only some traceless Hermitian matrices can be unitarily transformed into a purely imaginary form, so this is indeed a restriction on $\tilde{H}$ if $\tilde{H}^{(I)}$ is to exist at all. }
\end{thebibliography}

\end{document}